\documentclass[twocolumn,showpacs,preprintnumbers,amsmath,amssymb,prl]{revtex4}

\usepackage[english]{babel}
\usepackage{indentfirst}
\usepackage{graphicx}
\usepackage{amsmath}
\usepackage{amssymb}
\usepackage{amsthm}
\usepackage{mathrsfs}
\usepackage{lscape}
\usepackage{bm}
\usepackage{float}
\usepackage{longtable}
\usepackage{color}

\def \beq {\begin{equation}}
\def \eeq {\end{equation}}
\def \ba {\begin{eqnarray}}
\def \ea {\end{eqnarray}}
\newcommand{\upp}{\hspace{-0.2 pt}\uparrow}
\newcommand{\downn}{\hspace{-0.2 pt}\downarrow}

\def\ket#1{\left| #1\right>}

\begin{document}
\title{Direct Measurement of Quantum Dot Spin Dynamics \\ using Time-Resolved Resonance Fluorescence}
\author{C.-Y. Lu$^{1,\,2}$, Y. Zhao$^{1,\,3}$, A. N. Vamivakas$^{1}$, C. Matthiesen$^{1}$, S. Faelt$^{4}$, A. Badolato$^{5}$, M. Atat\"{u}re$^{1}$}
\affiliation{$^{1}$Cavendish Laboratory, University of Cambridge, JJ
Thomson Ave., Cambridge CB3 0HE, UK}
\affiliation{$^{2}$HFNL and Department of Modern Physics,
Univ. of Sci. Tech. of China, Hefei, 230026, China}
\affiliation{$^{3}$Physikalisches Institut,
Ruprecht-Karls-Universit\"{a}t Heidelberg, Philosophenweg 12, Heidelberg
69120, Germany} \affiliation{$^{4}$Sol Voltaics AB, Scheelevägen 17,
Ideon Science Park, 223 70 Lund, Sweden}
\affiliation{$^{5}$Department of Physics and Astronomy, University
of Rochester, Rochester, New York 14627, USA}

\vspace{-3.5cm}

\date{\today}
\begin{abstract}
We temporally resolve the resonance fluorescence from an electron spin confined to a single self-assembled quantum dot to measure directly the spin's optical initialization and natural relaxation timescales. Our measurements demonstrate that spin initialization occurs on the order of microseconds in the Faraday configuration when a laser resonantly drives the quantum dot transition. We show that the mechanism mediating the optically induced spin-flip changes from electron-nuclei interaction to hole-mixing interaction at 0.6 Tesla external magnetic field. Spin relaxation measurements result in times on the order of milliseconds and suggest that a $B^{-5}$ magnetic field dependence, due to spin-orbit coupling, is sustained all the way down to 2.2 Tesla.
\end{abstract}
 \maketitle


Single spins confined in semiconductor quantum dots (QDs) interact
with nearby charge, spin and phonon reservoirs in their solid state
environment. Signatures of these interactions are imprinted on the
spin's dynamics and elucidating the time scales relevant for these
couplings is not only interesting from the perspective of mesoscopic
physics, but is also important in assessing the potential of
a QD electron spin as a qubit in quantum information science \cite{Imamoglu99}. Driven
by these motivations a number of studies have begun to quantify both
spin relaxation and decoherence time scales
\cite{Kroutvar,Petta,Amasha,Elzerman,Johnson,Hanson,Reilly,yamamoto,greilich}. For spins confined in optically active semiconductor QDs, there is an additional timescale, namely the optically
induced spin-flip time $T_{P}$ - the time an optical field can
recycle a spin-selective dipole transition before a spin-flip event
is induced.

In this Letter, we present a direct $n$-shot measurement of spin
dynamics in a single self-assembled QD due to coupling to both an
optical field and the QD environment in the Faraday configuration.
We study the explicit dependence of the optically induced spin-flip rate on the properties of the optical field and identify the magnetic field value where the mechanism mediating the spin-flip
changes from ground-state mixing, due to electron-nuclei coupling,
to excited-state mixing, due to hole spins. We further demonstrate
that the natural spin relaxation rate, without the influence of an optical field, can vary more than two orders of
magnitude following the magnetic field dependence expected from
spin-orbit interaction inducing a ground-state spin admixture
\cite{Khaetskii,Woods}. Finally, we discuss briefly the prospect of
time resolved resonance fluorescence in the context of single-shot
read-out of spins in quantum dot systems.

The InAs/GaAs quantum dots studied in this work were grown by
molecular beam epitaxy (MBE) and embedded in a Schottky diode
heterostructure; the details of the device can be found in Ref.
\cite{vamivakas}. Such devices allow for deterministic charging of
QDs and we consider only the relevant ground and excited states for
a single electron charging under magnetic field in
the Faraday configuration, which can be understood by the 4-level
system illustrated in the inset of Fig. 1(b).  In this
representation the single electron ground states are spin down
$\ket{\downn}$ or spin up $\ket{\upp}$.  The two trion excited
states consist of an electron singlet and a single hole - depicted
as $\ket{\upp\downn\Uparrow}$ and $\ket{\downn\upp\Downarrow}$. The
magnetic field lifts the zero field degeneracy
between the two $X^{1-}$ transitions resulting in a blue (red) shift
for the $\ket{\upp}\leftrightarrow\ket{\upp\downn\Uparrow}$
($\ket{\downn}\leftrightarrow\ket{\downn\upp\Downarrow}$)
transition. These transitions are dipole allowed with a spontaneous
emission rate of $\sim2\pi \times 250$ MHz. The
$\ket{\downn}\leftrightarrow\ket{\upp\downn\Uparrow}$ transition is
normally forbidden due to the conservation of total angular
momentum, but weak interactions of QD spins with the environment
relax the optical selection rules and result in a weak
spontaneous emission rate $\gamma<<\Gamma$. As a result resonant optical excitation of
the blue (or red) transition can flip the spin of the ground state electron - a process we refer to as an optically induced spin flip
\cite{footnote1}. The rate $\xi_{\uparrow\downarrow}$ indicates
direct spin-flip transitions between the electronic ground states
$\ket{\upp}\leftrightarrow\ket{\downn}$ without the influence of an
optical field and can vary orders of magnitude as a function of
external parameters such as magnetic and electric field.

\begin{figure}[t]
\centering
  \includegraphics[width=0.5\textwidth]{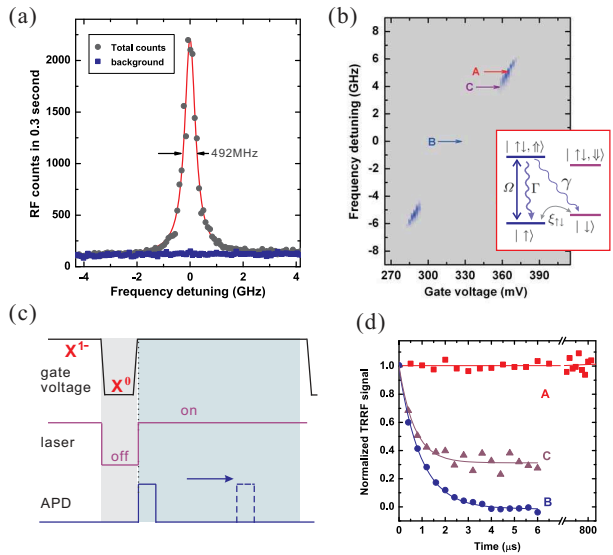}\\
  \caption{(Color online)(a) The integrated resonance fluorescence (grey circles) from the $X^{1-}$ transition as a function of laser detuning. The laser power is 0.2 times the spontaneous emission rate and there is no external magnetic field. Each data point corresponds to 300 ms of integration. The blue squares indicate the background level when the transition is far detuned via gate voltage.  (b) Two-dimensional DT map under 350 mT magnetic field. Inset: The reduced 4-level scheme for a deterministically charged QD. (c) One cycle of the protocol for measuring the optically induced spin-flip rate. (d) TRRF signal obtained at the three voltage-frequency conditions highlighted in panel b.
     }\label{}
\end{figure}


For the measurements reported here, the $X^{1-}$ transition is driven by a linearly
polarized, frequency and power stabilized, single mode laser. The
integrated resonance fluorescence (RF) \cite{vamivakas,Flagg} collected back through the
focusing objective passes through a second linear polarizer
(orthogonal to the laser polarization) prior to being sent to an
avalanche photodiode (APD).
In Fig. 1(a) the grey circles fit by the red curve present an
exemplary integrated RF spectrum as the laser frequency is swept
across the $X^{1-}$ transition. The blue squares in Fig. 1(a)
display the background, i.e. the same measurement when the QD
transition is Stark-shifted out of resonance. For this measurement,
the RF signal-to-background ratio is $18:1$ and the signal-to-noise
ratio is $160:1$ allowing for real-time transition monitoring at
higher bandwidth in comparison to other optical techniques such as
differential transmission.

To investigate the spin dynamics of the confined electron, we apply
a magnetic field in the Faraday configuration. Figure 1(b) displays
a 2-dimensional map of the differential transmission signal of the
blue trion transition at 350 mT magnetic field for the full single electron charging
plateau. In order to obtain a high precision measure of the
optically induced spin-flip timescale $T_{P}$, we resort to an
$n$-shot measurement of the integrated resonance fluorescence
spectrum. The details of one measurement protocol cycle are
presented in Fig. 1(c), where the first trace indicates the gate
voltage controlled charging state of the QD alternating between zero
and one excess electron. The second trace indicates the laser
amplitude, where the blue-shifted Zeeman transition is excited when
the laser is on. The third trace is the APD detection window. We
access the time dynamics of the integrated RF spectrum by recording
APD counts during a 0.1-2 $\mu$s time window which is scanned across
the on-window. We repeat this cycle $\sim$1x$10^5$ times for all
presented data. Background counts are subtracted for each cycle by far-detuning the transition.

Figure 1(d) presents time-resolved resonance fluorescence (TRRF)
measurements for 3 different combinations of laser frequency and
gate voltage highlighted as A, B and C in Fig. 1(b). The red curve
decorated with red squares is from location A at the center of the
cotunneling region of the charge-stability plateau.  In this region,
strong cotunneling with the Fermi sea of the back contact randomizes
the confined electron spin and mitigates any spin-pumping.
Consequently, there is no observable temporal dependence in the
emitted photon stream. The disappearance
of DT signal at location B was previously used to identify
efficient spin pumping resulting from optically induced back-action
and the absence of any mechanism leading to appreciable spin heating
during the time scales accessible to a DT measurement\cite{Atature06,Jan}. Nevertheless, the
transition still generates a photon stream, like a recycling
transition, until a single Stokes photon is emitted \cite{thomas}.
The blue curve decorated with blue circles in Fig. 1(d) presents
TRRF data from this location. Here, the transition initially
generates the same photon counts but the signal vanishes
exponentially within a few microseconds.  This timescale is to be
interpreted in two ways: From the perspective of state preparation,
it takes a few $\mu$s to initialize the electron spin. Alternatively, the transition can be recycled for a few $\mu$s
before laser induces an unwanted spin-flip
event. Identifying the physical mechanisms that lead to this
observation is thus of interest from both perspectives. Finally, the
TRRF signal at location C displays the intermediate dynamics: The
initial exponential decay due to optical spin pumping is still
present, but the signal saturates at a constant value determined by
the ratio of spin pumping and cotunneling rates. In this case, the cotunneling rate is $2\pi\times37$ KHz.

\begin{figure}[t]
\centering
  \includegraphics[width=0.5\textwidth]{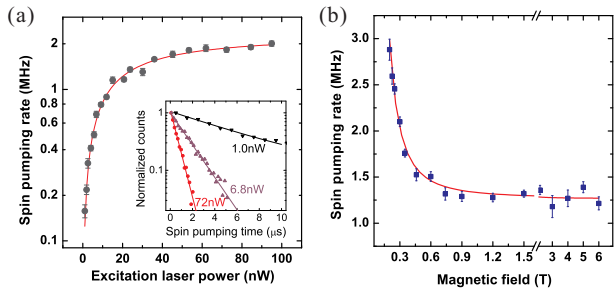}\\
  \caption{(Color online) (a) The extracted spin-flip rate from TRRF measurements for a range of laser powers when the laser is resonant with the $X^{1-}$ transition. The saturation power is 18.7 nW. inset: A log-plot of TRRF signal for 3 laser powers. (b) The magnetic field dependence of the spin-flip rate for a fixed laser power of 60 nW. The lowest magnetic field value of 200 mT is selected to ensure the electronic ground states are split by $\sim$1.5 GHz which is 3 times the transition linewidth.
     }\label{}
\end{figure}

Next, we analyze how the excitation Rabi frequency influences $T_{P}$ at a fixed magnetic field.  The state mixing for
fixed magnetic fields yields a branching ratio $\eta$,  defined as
$\gamma/(\Gamma+\gamma)$ \cite{Jan}, which quantifies the number of
photons cycled by the transition before the electron flips its spin.
With a fixed branching ratio, the optically induced spin-flip rate is
determined by the population in the excited state
$\ket{\upp\downn\Uparrow}$. Consequently, the spin pumping rate is
expected to increase with excitation Rabi frequency until it reaches
a saturation value.
The inset of Fig. 2(a) displays a log-plot of the TRRF signal obtained for three
excitation powers for the gate voltage and laser frequency
combination labeled B in Fig. 1(b). The black squares in Fig. 2(a)
are the extracted optical spin pumping rates per excitation power.
In the limit where the Rabi frequency is much larger than the
spontaneous emission rate, the spin flip rate saturates at a rate of $2\pi\times200$KHz.

To elucidate the physical mechanisms which mediate the optical induced spin-flip, we study the magnetic field dependence of $T_{P}$. The laser
power is set well above the saturation power and the laser frequency
and gate voltage are fixed at the position equivalent to B in Fig.
1(b) for each magnetic field value. Figure 2(b) presents the
magnetic field dependence of $T_{P}$. Each data point is extracted from TRRF
measurements that map out the detuning dependence of the spin pumping rate (see supplementary information) in order to ensure that all
measurements are obtained on resonance with the transition in the
spin-pumping region. The fitting curve includes the functional
dependence on magnetic field of two dominant mechanisms that mix
the spin states coherently, namely the hyperfine interaction and
the heavy-light hole mixing. In the low magnetic
field limit, the hyperfine interaction efficiently mediates the
spin-flip process and results in a quadratic variation of the
spin-flip time with applied external magnetic field according to
$(B_{N}/B_{ext})^2$. Whereas, for magnetic fields beyond $0.6$ T,
hole mixing mediates the spin-flip process and is independent of
external magnetic field. The theoretical curve is obtained using an
RMS nuclear field of 15 mT. The corresponding heavy-light hole mixing strength is $\left|\epsilon_{hl}\right|=2.8$\% for the QD presented
here\cite{footnote2p5}, which is within the estimated range based on previous reports
using differential transmission measurements \cite{Jan,footnote3}.
We do note that the value of hole-mixing strength will vary among
QDs due to the shape anisotropy and the large variation of hole-spin
g-factor.

\begin{figure}[t]
\centering
  \includegraphics[width=0.5\textwidth]{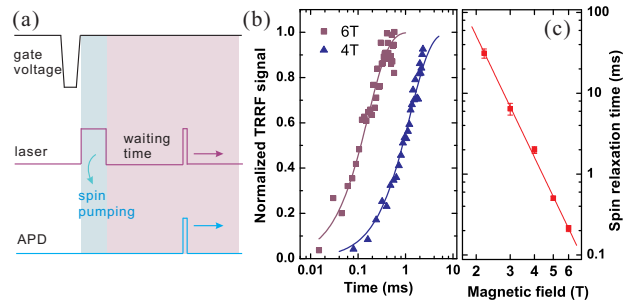}\\
  \caption{(Color online) (a) One cycle of the protocol for measuring the spin relaxation
  timescale. We highlight that at the end of the on-window there is an additional
pulse sent to the QD gate (not shown). This artificially but
deterministically recycles the QD electron spin and recovers the
expected steady-state signal value as determined by the Boltzmann
statistics and allows a higher accuracy in theoretical fits.
  (b) Exemplary TRRF measurements of the spin relaxation timescale for two magnetic field values.
  (c) The extracted spin relaxation time $T_{\downn \upp}$ (red squares) as a function
  of magnetic field. The red curve is the best fit $B^{-5}$ dependence.}    \label{}
\end{figure}

We now present measurements of the natural spin dynamics directly between
the two ground states.
The protocol for measuring the spin
relaxation is illustrated in Fig. 3(a).  The laser is turned on for 50$\mu$s at the beginning of each cycle,
to ensure spin initialization into $\ket{\downn}$, then the electron is left in the dark for a waiting
time spanning 0-20 msec.  The laser is then turned back on for
5$\mu$s coinciding in time with the detection window. The set
of time traces in Fig. 3(b) shows the measured signal recovery for
two magnetic field values. Initially, the electron still
resides in the dark spin down state $\ket{\downn}$ and no photon
scattering occurs.  As time progresses, and the probability that the
electron has flipped its spin orientation increases, the probability
to scatter photons also increases. Using the functional dependence
of $\rho_{\upp \upp} \sim a(1-e^{-t/T_{eff}})$ we extract a
corresponding \emph{effective} spin-flip time per magnetic field.
Here, $T_{eff} = T_{\upp \downn} T_{\downn \upp} / (T_{\upp
\downn} + T_{\downn \upp})$ following $(T_{\downn
\upp} / T_{\upp \downn}) = e^{(-g_{e}\mu_{B}B/k_{B}T)}$, where
$g_{e}$ is the electronic g-factor. The highest measured $T_{eff}$
of 17.3 ms at 2.2 Tesla corresponds to a $T_{\downn \upp}$ time of
31.3 ms. The red curve in Fig. 3(c) displays a close agreement with the $B^{-5}$
magnetic field dependence expected for spin relaxation due to
single-phonon assisted spin orbit coupling \cite{power-law}. This
power dependence indicates that the spin-orbit interaction
inducing an admixture of electronic ground and excited states is the dominant mechanism even in the
case of $k_{B}T > g_{e}\mu_{B}B$. This
mechanism depends strongly on the electron orbital wavefunctions and gate-voltage
control of the confinement potential can modify the spin relaxation
rate in electrostatic QDs \cite{Amasha}. In self-assembled QDs, a
similar correlation between the magnitude and orientation of the X-Y
splitting of the neutral exciton and the spin relaxation rate can be
investigated using this technique. The X-Y splitting for the QD
studied here is 22 $\mu$eV, twice the mean value of a QD ensemble \cite{footnote4}.

\begin{figure}[b]
\centering
  \includegraphics[width=0.4\textwidth]{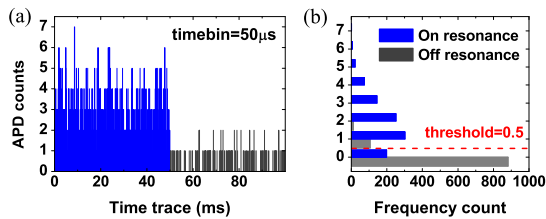}\\
  \caption{(Color online) (a) Real-time monitoring of the photon stream scattered from the $X^{1-}$ transition
  at zero magnetic field (the first 50 ms) and the background counts (the second 50 ms) with a
  time-bin of 50 $\mu$s (left panel), and (b) the corresponding histograms.}
\end{figure}

While $n$-shot measurements were employed to record spin
initialization and relaxation processes, it is of great interest to
quantify the shortest time needed to identify with sufficient
fidelity, whether we are probing on resonance or off resonance. In
Fig. 4 we present 100 ms worth of real time RF counts with a
50$\mu$s time bin; for data on 10$\mu$s and 30$\mu$s time bins we refer to the supplementary information. The first 50 ms time trace is obtained when the
trion transition is resonant with the excitation laser at $\sim$10
times the saturation power. The second 50 ms part is obtained when
the transition is far off resonance with the laser dictating the
overall background level. The magnetic field is set to zero in order
not to obscure the measurements with spin pumping. The read-out error is defined
as $\epsilon = \frac{1}{2}(\epsilon_{on} + \epsilon_{off})$
\cite{ion}, where $\epsilon_{on}$ ($\epsilon_{off}$) is the fraction
of detection attempts where the transition is on (off), but declared
to be off (on) since the count is below (above) the set threshold.
With a threshold of 0.5 we deduce measurement fidelities
($1-\epsilon$) of 0.63, 0.77, and 0.84 for the 10, 30, and 50 $\mu$s
time bins, respectively. These numbers are satisfactory when
compared to spin relaxation timescales and single-shot read-out is in
principle possible with sufficient margin with respect to all spin relaxation times of Fig. 3(c). However, for finite magnetic fields the optically induced spin-flip
time sets the natural limit for (non-destructive) readout
in trionic transition of a single QD configuration, necessitating a
measurement time much shorter that 1 $\mu$s. Alternatively, a
different QD system demonstrating similar optical measurement and
spin relaxation timescales, but not limited by such a short
optically induced spin-flip time can be utilized for this purpose. One strong
candidate for using TRRF to reveal spin quantum jumps is tunnel-coupled quantum dot pairs \cite{lucio,kim}, where one QD confines a
single excess electron and the other QD is in neutral charge
configuration. In such a system, the frequency selective probing of
the $\ket{\downn}\rightarrow\ket{\downn\downn\Uparrow}$ transition
(spectrally shifted from the
$\ket{\upp}\rightarrow\ket{\upp\downn\Uparrow}$ transition) with our
TRRF technique is expected to yield real-time dynamics of electron
spin based on the measured timescales reported here.

In summary, we have shown that the time-resolved resonance fluorescence
technique introduced here allows for accurate and direct measurement
of parameters essential to QD electron spin dynamics, namely
excitation spin-flip scattering and spin relaxation timescales. In addition
to demonstrating near-background free QD emission at the $\Omega
\approx \Gamma$ resonant excitation regime, we have also shown explicitly the crossover, from hyperfine to hole-mixing, in the mechanism that mediates the optically induced flip of the spin.  We have further observed a
single magnetic field dependence of the spin relaxation rate down to
2.2 Tesla which indicates that the dominant reason for spin
relaxation is single-phonon assisted spin-orbit coupling mechanism.
Our results indicate that sub-microsecond real-time resolution is
necessary for the single-shot measurement of spins using resonant
excitation for a single QD configuration, which we do not
demonstrate in this work. A signal to noise improvement of 3-10, or
working with coupled QD systems with our signal-to-noise values would
in principle allow this type of direct monitoring. Another natural extension
of this work will be looking for correlations between QD anisotropy and
electron spin relaxation rates.


This work was supported by EPSRC grant no. EP/G000883/1, QIP IRC, and Univ. of
Cambridge. The authors acknowledge fruitful discussions
with A. Imamoglu, J. Taylor, L. Jiang and R.B. Liu.


\begin{thebibliography}{99}
\bibitem{footnote1} The relaxation of the selection rules on the $\ket{\downn}\leftrightarrow\ket{\upp\downn\Uparrow}$ transition results in both spontaneous Raman scattering from $\ket{\upp}\leftrightarrow\ket{\downn}$ and spontaneous emission on the $\ket{\upp\downn\Uparrow}\leftrightarrow\ket{\downn}$ channel when driving $\ket{\upp}\leftrightarrow\ket{\upp\downn\Uparrow}$.  Both of these processes flip the ground state electron spin and we refer to either of these processes as an optically induced spin-flip.
\bibitem{footnote2p5} Here we use the notation, $\ket{\Uparrow}=\ket{+3/2}+\epsilon_{+}\ket{+1/2}+\epsilon_{-}\ket{-1/2}$ and $|\epsilon_{hl}|=\left[|\epsilon_{+}|^2+|\epsilon_{-}|^2\right]^{1/2}$.
\bibitem{footnote3} We note that beyond 2 Tesla external magnetic field, TRRF measurements similar to Fig. 2(b) show a dragging of the transition due to dynamical nuclear spin polarization \cite{Latta09, Xu09}. For the $T_{BA}$ measurements we operate at nearly zero net nuclear spin polarization and all measurements are performed with constant laser frequency and gate voltage.
\bibitem{power-law} A deviation from the $B^{-5}$ to the thermal energy corrected $k_{B}T B^{-4}$ dependence \cite{Khaetskii} was not observed for the magnetic field range of our experiments.
\bibitem{footnote4} The deviation from the average value (obtained from measurements on 20 QDs) is likely to result in shorter $T_{1}$ times due to anisotropy enhanced spin-orbit coupling than similar measurements on a QD ensemble.
\end{thebibliography}
\end{document}